\documentclass[%
 reprint,
 amsmath,amssymb,
 aps,
]{revtex4-2}

\usepackage{graphicx}
\usepackage{dcolumn}
\usepackage{bm}
\usepackage{xcolor}

\begin{document}

\preprint{APS/123-QED}

\title{Comment on ``In search of the electron-phonon contribution to total 
energy"}

\author{Shilpa Paul}%
\affiliation{%
	Department of Metallurgical Engineering and Materials Science, Indian Institute of Technology Bombay, Mumbai 400076, India
}%

\author{Amrita Bhattacharya}%
\affiliation{%
	Department of Metallurgical Engineering and Materials Science, Indian Institute of Technology Bombay, Mumbai 400076, India
}%
\author{Tiramkudlu R. S. Prasanna}%
\email{prasanna@iitb.ac.in}
\affiliation{%
	Department of Metallurgical Engineering and Materials Science, Indian Institute of Technology Bombay, Mumbai 400076, India
}%

\date{\today}

\begin{abstract}
 	Recently, Ponc{\'e} and Gonze (PRX Energy $\mathbf{5}$, 043004 (2026), 
 	arXiv:2512.04897) have proposed a new theoretical formalism to account for 
 	the electron-phonon contribution to 
 	the \textit{ab initio} total energy. In particular, a key result is stated 
 	as ``We clarify that the expression proposed by Allen in 2020, which he 
 	called the electron-phonon contribution to the total energy, is in fact 
 	part of the second-order phonon contribution and should not be added 
 	independently." We have analysed the implications of the above result for 
 	Boron Nitride polymorphs. Our results show that the total energy in the 
 	effective Hamiltonian formalism  gives a better match with the experimental 
 	results, rather than the new formalism. This motivated us  to closely study 
 	the new formalism. The key result in the new formalism has been derived for 
 	the specific case of the adiabatic approximation. We show that this result 
 	is not valid for the general case of the  non-adiabatic approximation which 
 	is to be used for all IR-active materials. Further, relying on Allen's 
 	objections, we show that the above result is theoretically unjustifiable 
 	even for the adiabatic approximation, on thermodynamic grounds. For all the 
 	above reasons the effective Hamiltonian formalism must be preferred in 
 	total energy studies.

\end{abstract}

\maketitle


	Until recently, the \textit{ab initio} total energy consisted of the 
	electronic energy, computed using the Density Functional Theory (DFT) and 
	the zero-point vibrational energy (ZPVE), computed from (computed from the 
	Density functional perturbation theory (DFPT) or Finite difference method 
	(FDM) \cite{Kawanishi2016, Scalise2019, Ramakers, Cazorla2019, Van2007, 
	Bhat2018}. In addition, the dispersion correction or the van der 
	Waals (vdW) contribution was also included \cite{Kawanishi2016, 
	Scalise2019, Ramakers, Cazorla2019}). 

	We have recently proposed a new contribution to the \textit{ab initio} 
	total energy, due to electron-phonon interactions \cite{Varma2022}. 
	``Stimulated" by our work, Allen has derived the expression for the EPI 
	contribution to the total energy as a part of the effective Hamiltonian 
	\cite{Allen2022erratum}. Thus, in the effective Hamiltonian 
	formalism, the \textit{ab initio} total energy is given by DFT (including 
	vdW) energy, ZPVE (phonon) and EPI energy  \cite{Varma2022, 
	Allen2022erratum}. 

	Very recently,  Ponc{\'e} and Gonze \cite{Ponce2026PRXEnergy} have proposed 
	a new formalism for the total energy. In this formalism, the EPI 
	contribution derived by Allen \cite{Allen2022erratum} is part of the phonon 
	energy term, ZPVE, and is not a separate term \cite{Ponce2026PRXEnergy}. 
	The EPI makes an additional contribution only in the fourth-order term. 
	Thus, upto second-order, the total energy in the new formalism 
	\cite{Ponce2026PRXEnergy} is identical to the \textit{ab initio} total 
	energy used earlier \cite{Kawanishi2016, Scalise2019, Ramakers, 
	Cazorla2019, Van2007, Bhat2018}, before the introduction of the EPI term 
	\cite{Varma2022, Allen2022erratum}.

	Hence, the key difference in the two formalisms is the presence/absence of 
	a separate EPI term in the total energy expression. For this reason, we 
	have studied the energy differences between Boron Nitride polymorphs using 
	both formalisms and compared them with experimental results 
	\cite{Paul2022}. Our analysis reveals serious concerns with the 
	total energy in the new formalism. In this comment, we first discuss our 
	theoretical concerns about the new formalism. Later, we compare the total 
	energy differences computed in both formalisms with the experimental 
	results for BN polymorphs. For all the above reasons, we suggest that  the 
	total energy in the effective Hamiltonian formalism, rather than the new 
	formalism, must be preferred.

	\section{Theoretical concerns}

\subsection*{Point-1}
\subsubsection*{The EPI contribution in the non-adiabatic approximation is not 
a part of ZPVE}
	We begin with the key result of Ponc{\'e} and Gonze 
	\cite{Ponce2026PRXEnergy}, \textit{viz.} Eq.112,  given as
\begin{equation}
     E_{\textrm{ZPVE}}^{\textrm{ph}} = 2 E^{\textrm{Allen-EPI}} +  
     E^{\textrm{ph,other}}
	\label{Eph-total}
	\end{equation}
	where we have added the notations EPI and ZPVE to improve clarity. It shows 
	that the EPI contribution derived by Allen \cite{Allen2022erratum} is part 
	of the phonon energy, ZPVE.

	We stress that Eq. \ref{Eph-total} has been derived for the special case 
	of the adiabatic approximation. (The starting point is the authors' Eq.108, 
	\textit{viz.} the expression for the EPI contribution to each eigenstate, 
	$\Delta \epsilon_{n\textbf{k}}$, which is obtained from Eq.1 
	and Eq.2, from the section titled ``The AHC theory \textit{within 
	the adiabatic harmonic approximation}" of Ref. \cite{Ponce2015}). The 
	adiabatic approximation is valid for IR-inactive materials, i.e. to C, Si 
	and Ge polymorphs. However, the vast majority of semiconductors and 
	insulators are IR-active and the non-adiabatic approximation must be used 
	\cite{Ponce2015, Miglio2020}. This is also	explicitly stated by the 
	authors \cite{Ponce2026PRXEnergy} as ``We also 	
	remark that the eigenvalue renormalization in Eq. (104) is expressed using 	
	the adiabatic approximation, but the nonadiabatic version of $ \Delta 
	\epsilon_{n\textbf{k}}$ should be used for infrared-active materials." 

	\textit{However, it is crucial to note that it is the non-adiabatic 
	version of Allen's equation for the EPI contribution to the total energy 
	that must be used.} This is given by \cite{Allen2022erratum, Allen2023}
	\begin{equation}
    \begin{split}
		\Delta E_{\textrm{EP}} (V,0) &= \sum_k \Bigg[\langle k 
		|V^{(2)}|k\rangle \\&+ \sum_Q \frac{|\langle k 
			|V^{(1)}|k+Q\rangle|^2}{\epsilon_k - 
			\epsilon_{k+Q}-\hbar\omega_{-Q}}(1-f_{k+Q}) \Bigg]\;f_k
		\label{EPI-zero-K-Allen}
    \end{split}
	\end{equation}
	We note that Eq. \ref{EPI-zero-K-Allen} includes the phonon energy term in 
	the denominator, which is not present in Eq.4 of Ref. 
	\cite{Allen2022erratum}. Allen \cite{Allen2023} has clarified that it was 
	an oversight and provided the above Eq. \ref{EPI-zero-K-Allen} and, in 
	addition, also provided a more explicit expression for the second term as 
	\cite{Allen2023}
	\begin{equation}
		\Delta 
		E_{\textrm{tot}}^{\textrm{FM,unocc}} = \\2 \sum_{\vec{k}n}^{occ} 
		\sum_{\vec{k}+\vec{Q}n'}^{unocc}\sum_{j=1}^{3n_{at}}
		\frac{|M_{\vec{k}n,\vec{k}+\vec{Q}n',j}|^2}{\epsilon_{\vec{k}n} -
			\epsilon_{\vec{k}+\vec{Q}n'}-\hbar\omega_{\textbf qj}}
		\label{EPI-FM-unocc-Allen}
	\end{equation}
	``where the 2 is for spin and $n_{at}$ is the number of atoms in the 
	supercell used." This is the Fan-Migdal contribution.

	Eq. \ref{EPI-FM-unocc-Allen} is identical to earlier 
	expressions of Fr{\"o}hlich \cite{Frohlich1950theory} and Fan 
	\cite{Fan1951}. For example, Eq.2.11  of Fr{\"o}hlich 
	\cite{Frohlich1950theory} gives the FM contribution to the total energy as
	\begin{equation}
		E = - 2 \sum_{\mathbf{k}} \sum_{\mathbf{w}} 
		\frac{|M_{w}|^2 f_\mathbf{k} (1-f_\mathbf{q})}{\epsilon_{q} -
			\epsilon_{k}+\hbar s w}
		\label{Frohlich 2.11}
	\end{equation}
	Further, Eq.4 of Fan \cite{Fan1951} gives the FM contribution to the total 
	energy as
\begin{equation}
	\Delta E = \sum_{\mathbf{k},\sigma} \sum_{\mathbf{q},\sigma'} \frac{
		\left| \left\langle \psi(\mathbf{k}\pm\mathbf{q},\sigma') 
		\chi(n_{\mathbf{q}}\mp 1)
		\middle| \Delta U \middle| \psi(\mathbf{k},\sigma) \chi(n_{\mathbf{q}})
		\right\rangle \right|^2 }{ \epsilon(\mathbf{k},\sigma) -\left[
		\epsilon(\mathbf{k}\pm\mathbf{q},\sigma') \mp \hbar\omega_{\mathbf{q}}
		\right]}
		\label{Fan Eq.4}
\end{equation}
	The above equations, Eq. \ref{EPI-zero-K-Allen} to Eq. \ref{Fan Eq.4},  all 
	contain the phonon energy term in the denominator. 	
	\textit{Hence, it is clear that Allen's equation  is for the general case 
	of the non-adiabatic approximation. In particular, the FM term, which is 
	$\Delta E_{\textrm{tot}}^{\textrm{FM,unocc}}$, is consistent with the 
	earliest studies of Fr{\"o}hlich \cite{Frohlich1950theory} and  Fan 		
	\cite{Fan1951}.}	

	Importantly, the FM contribution to any eigenstate can be expressed 
	as a sum of occupied and unoccupied partial-FM terms, $\Delta 
	\epsilon_{n\textbf{k}}^{\textrm{FM,occ}}$ and $\Delta 
	\epsilon_{n\textbf{k}}^{\textrm{FM,unocc}}$ \cite{Nery2018}. 

In Eq. \ref{EPI-zero-K-Allen}, the summation of the FM term is 
only over the unoccupied states due to the $(1-f_{k+Q})$ factor, as also seen 
from Eq. \ref{EPI-FM-unocc-Allen} (and Eq.4 and Eq.5 above 
\cite{Frohlich1950theory, Fan1951}). The FM term can be rewritten 
\cite{Paul2026} as the full-FM term minus the FM-occupied term. 
Combining the full FM term with the first DW term gives the EPI contribution to 
the band-structure energy. The EPI contribution to the total energy is given by
	\begin{equation}
		\begin{split}
			\Delta E_{\textrm{tot}}^{\textrm{ep}}  
			&= \Delta E^{\textrm{ep}}_{\textrm{bs}} - \Delta 
			E_{\textrm{tot}}^{\textrm{FM,occ}}
			\label{EPI-zero-K-Allen-rewritten}
		\end{split}
	\end{equation}
	where the first term is the EPI contribution to the band-structure energy 
	given by
	\begin{equation}
		\Delta E^{\textrm{ep}}_{\textrm{bs}} =  
		\sum^{\textrm{occ}}_{n,\textbf{k}}  \Delta \epsilon_{n\textbf{k}}
		\label{EPI-band-struc-energy}
	\end{equation}
	and is identical to Eq. 107 of Ponc{\'e} and Gonze 
	\cite{Ponce2026PRXEnergy}. The second term is the partial-FM-occupied 
	contribution to the 
	total energy and is given by
	\begin{equation}
\begin{split}
 \Delta E_{\textrm{tot}}^{\textrm{FM,occ}} & =  
 \sum_{\mathbf{k},\mathbf{q}}^{\textrm{BZ}}
\sum_{\nu nm}^{occ}
\frac{\left|g_{mn\nu} (\mathbf{k}, 
\mathbf{q})\right|^{2}}{\epsilon_{n\mathbf{k}} - 
\epsilon_{m\mathbf{k}+\mathbf{q}} + \hbar\omega_{\textbf q\nu} } 
			\label{EPI-FM-occ-tot}
		\end{split}
	\end{equation}
In the adiabatic approximation, the phonon energy term ($\hbar\omega_{\textbf 
q\nu}$) is set = 0 and the denominator in Eq. \ref{EPI-FM-occ-tot} becomes 
$(\epsilon_{n\mathbf{k}} - \epsilon_{m\mathbf{k}+\mathbf{q}})$. In this case, 
Ponc{\'e} and Gonze \cite{Ponce2026PRXEnergy} have pointed out that because the 
sign of the denominator changes upon exchange of ${n\mathbf{k}}$ and 
${m\mathbf{k}+\mathbf{q}}$,  Eq. \ref{EPI-FM-occ-tot} is identically zero.  Due 
to this result, the EPI contribution to the total energy in the adiabatic 
approximation becomes
	\begin{equation}
		\begin{split}
			\Delta E_{\textrm{tot,adia}}^{\textrm{ep}}  
			&= \Delta E^{\textrm{ep}}_{\textrm{bs,adia}} 			
			\label{EPI-zero-K-Allen-adia}
		\end{split}
	\end{equation}
Eq. \ref{EPI-zero-K-Allen-adia} is identical to the authors' Eq.107 
($E_{\textrm{adia}}^{\textrm{Allen-EPI}}$). Further, the authors 
\cite{Ponce2026PRXEnergy} have shown this term to be a part of the phonon 
energy in  Eq. \ref{Eph-total} (authors' Eq.112).

\textit{It is crucial to note that Eq. \ref{EPI-zero-K-Allen-adia} only holds 
in the adiabatic approximation where the phonon energy in Eq. 
\ref{EPI-FM-occ-tot} is set to zero}. 

However, in the non-adiabatic approximation, due to the presence of the 
$\hbar\omega_{\textbf q\nu}$ term, Eq. \ref{EPI-FM-occ-tot} is non-zero. The 
EPI contribution to the total energy in the non-adiabatic contribution is, 
hence, given by
\begin{equation}
		\begin{split}
			\Delta E_{\textrm{tot,non-adia}}^{\textrm{ep}}  
			&= \Delta E^{\textrm{ep}}_{\textrm{bs,non-adia}} - \Delta 
			E_{\textrm{tot,non-adia}}^{\textrm{FM,occ}}
			\label{EPI-zero-K-Allen-non-adia}
		\end{split}
	\end{equation}
We recall that the derivation of Eq. \ref{Eph-total} (authors' Eq.112) begins 
with the authors' Eq.108, the expression for  $\Delta 
\epsilon_{n\textbf{k}}$. Unlike the first term in Eq. 
\ref{EPI-zero-K-Allen-non-adia}, the second term  cannot be derived by starting 
from  $\Delta \epsilon_{n\textbf{k}}$. (The second term is obtained from the 
partial-FM term, $\Delta \epsilon_{n\textbf{k}}^{\textrm{FM,occ}}$.) Therefore, 
because of the presence of $\Delta E_{\textrm{tot,non-adia}}^{\textrm{FM,occ}}$ 
term in Eq. \ref{EPI-zero-K-Allen-non-adia}, \textit{the EPI contribution to 
the total energy is not a part ZPVE}. Hence, the key result, Eq. 
\ref{Eph-total} (authors' Eq.112), is not valid for the non-adiabatic 
approximation.

\subsection*{Point-2}
	\subsubsection*{The EPI contribution to the band-structure energy in the 
	non-adiabatic approximation is not a part of ZPVE}	

	The non-adiabatic effects on the renormalization of electron and phonon 
	energies are drastically different. 

	The phonon energy is given as the sum of the adiabatic and 
	non-adiabatic parts in Eq.125 of Giustino \cite{Giustino2017} as
\begin{equation}
\Pi(\omega) = \Pi^{\mathrm{A}} + \Pi^{\mathrm{NA}}(\omega)
\label{Giustino-phonon self energy}
\end{equation}
The non-adiabatic contribution is given by Eq. 145 of Giustino 
\cite{Giustino2017} as 
 \begin{equation}
      \begin{split}
      	\hbar&\Pi^{\mathrm{NA}}_{\mathbf{q}\nu\nu'}(\omega)
      	= 2\sum_{mn} \int \frac{d\mathbf{k}}{\Omega_{\mathrm{BZ}}}
      	g^{\mathrm{b}}_{mn\nu}(\mathbf{k},\mathbf{q})
      	g^{*}_{mn\nu'}(\mathbf{k},\mathbf{q}) \\
      	& \times \left[
      	\frac{f_{m\mathbf{k}+\mathbf{q}}-f_{n\mathbf{k}}}
      	{\epsilon_{m\mathbf{k}+\mathbf{q}}-\epsilon_{n\mathbf{k}}
      		-\hbar(\omega+i\eta)} \right. \left. - \frac{f_{m\mathbf{k}+\mathbf{q}}-f_{n\mathbf{k}}}
      	{\epsilon_{m\mathbf{k}+\mathbf{q}}-\epsilon_{n\mathbf{k}}}
      	\right]
      \label{Giustino-phonon energy_NA}
      \end{split}
      \end{equation}

	 In Eq. \ref{Giustino-phonon energy_NA} above, the Fermi-Dirac 
	 factors ensure that the mediating phonon connects the occupied and 
	 unoccupied states in $(\epsilon_{m\mathbf{k}+\mathbf{q}} - 
	 \epsilon_{n\mathbf{k}})$ \cite{Calandra2010,Giustino2017}. For large 
	 band-gap materials, with  $|\epsilon_{m\mathbf{k}+\mathbf{q}} - 
	 \epsilon_{n\mathbf{k}}| \gg  \hbar\omega_{\textbf q\nu}$, the contribution 
	 to phonon renormalization is negligible \cite{Calandra2010,  
	 Giustino2017}. The  non-adiabatic effects are  relevant  only for metals 
	 and small band-gap  materials \cite{Calandra2010,  Giustino2017}. Hence, 
	 in the case of  semiconductors and insulators, the  ZPVE computed in the 
	 adiabatic  approximation is valid  in the  non-adiabatic approximation as 
	 well  \cite{Calandra2010,Giustino2017}.

	In contrast, for electrons, Eq.157 of 	Giustino \cite{Giustino2017} give 
	the expression 
	for the Fan-Migdal term as 
  \begin{equation}
  	\begin{split}
  &\Sigma^{\mathrm{FM}}_{nn'\mathbf{k}}(\omega)
  = \frac{1}{\hbar} \sum_{m\nu} \int 
  \frac{d\mathbf{q}}{{\Omega_{\mathrm{BZ}}}}
   g^{*}_{mn\nu}(\mathbf{k},\mathbf{q}) g_{mn'\nu}(\mathbf{k},\mathbf{q}) \\
  &\times \left[ 
  \frac{1-f_{m\mathbf{k}+\mathbf{q}}}
  {\omega-\epsilon_{m\mathbf{k}+\mathbf{q}}/\hbar-\omega_{\mathbf{q}\nu}+i\eta}
   \right. \left.+ \frac{f_{m\mathbf{k}+\mathbf{q}}} 
  {\omega-\epsilon_{m\mathbf{k}+\mathbf{q}} 
  /\hbar+\omega_{\mathbf{q}\nu}-i\eta} \right]
  \label{Giustino-FM self energy}
 	 \end{split}
  	\end{equation}
	In Eq. \ref{Giustino-FM self energy}, the phonon energy term is set to zero 
	in the adiabatic approximation. Whereas, in the non-adiabatic 
	approximation, the phonon energy term is considered. Importantly, in the 
	summation over electronic energy bands, there is no restriction (unlike the 
	case for phonons) and the 
	summation is over all bands. Hence, the difference between the adiabatic 
	and non-adiabatic FM contribution will be a finite quantity. Therefore the 
	EPI contribution to individual eigenstates, $\Delta 
	\epsilon_{n\textbf{k}}$, and hence, to the band-structure energy, will 
	be different. In short, there is a finite difference between $\Delta 
	E^{\textrm{ep}}_{\textrm{bs,non-adia}}$ and $\Delta 
	E^{\textrm{ep}}_{\textrm{bs,adia}}$. 

	Therefore, in Eq. \ref{Eph-total} (authors' Eq.112), for 
	$E^{\textrm{Allen-EPI}}$, if $\Delta E^{\textrm{ep}}_{\textrm{bs,adia}}$ is 
	replaced by $\Delta	E^{\textrm{ep}}_{\textrm{bs,non-adia}}$ and ZPVE is a 
	constant, it implies that the $E^{\textrm{ph,other}}$ term varies with the 
	approximation used. This is unacceptable as it is a constant. This is 
	seen more clearly from the authors's  Eq. 93, which shows that the 
	expression for 	$E^{\textrm{other}}$ depends only on the DFT results 
	\cite{Ponce2026PRXEnergy}.  $E^{\textrm{ph,other}}$ is obtained from the 
	authors' Eq.113 as
\begin{equation}
\begin{split}
&E^{\mathrm{ph,other}}
=
\sum_{\nu}
\int_{\mathrm{BZ}}
\frac{\mathrm{d\mathbf{q}}}{\Omega_{\mathrm{BZ}}}
\frac{1}{2\omega_{\nu\mathbf{q}}} \\
&\times
\sum_{\kappa \alpha\kappa'\beta l}
\frac{\partial^2 E^{\mathrm{other}}}
{\partial R_{0\kappa\alpha}\,\partial R_{l\kappa'\beta}}
e^{i\mathbf{q}\cdot\mathbf{R}_l}
U_{\nu\kappa\alpha\mathbf{q}}^{*}
U_{\nu\kappa'\beta\mathbf{q}}
\end{split}
\label{PonceEph-other}
\end{equation}
	Beacause $E^{\textrm{other}}$ is obtained from DFT, it follows from Eq. 
	\ref{PonceEph-other} (authors' Eq.113) that $E^{\textrm{ph,other}}$  will 
	remain the same irrespective of the approximation used.	Hence, in Eq. 
	\ref{Eph-total} (authors' Eq.112), only $\Delta 
	E^{\textrm{ep}}_{\textrm{bs,adia}}$ can be used because replacing it 
	with $\Delta	E^{\textrm{ep}}_{\textrm{bs,non-adia}}$ leads to a 
	different value of $E^{\textrm{ph,other}}$. Clearly, the replacement is not 
	theoretically justifiable

	Thus, in the non-adiabatic approximation, \textit{ both the terms} in the 
	EPI	contribution to the total energy,  Eq. \ref{EPI-zero-K-Allen-non-adia}, 
	are not part of ZPVE  in Eq. \ref{Eph-total} (authors' 
	Eq.112).

	\subsection*{Point-3}
	\subsubsection*{Computational results in support of Point-2}

	The non-adiabatic approximation is the general case 
	that is valid for all materials, IR-active and IR-inactive. However, the 
	specific case of the adiabatic approximation is valid only for IR-inactive 
	materials, because the issues related to divergences are not present 
	\cite{Ponce2015, Miglio2020}. 

Carbon-diamond (C-dia) is IR-inactive \cite{Ponce2015} and hence, the EPI 	
	contribution is valid in both approximations. This is a good 
	example to highlight the claim in Point-2.

	We compute $E^{\textrm{ph,other}}$ by rewriting the authors' Eq.112 (Eq. 
	\ref{Eph-total}) as 
	\begin{equation}
     E^{\textrm{ph,other}} = E^{\textrm{ph}}_{ZPVE} - 2  
     E^{\textrm{Allen-EPI}}   
	\label{Eph-other}
	\end{equation}
	Indeed, using this relation, the authors \cite{Ponce2026PRXEnergy} have 
	computed $E^{\textrm{ph,other}}$ in their Table II.

	We begin with the zero-point renormalization (ZPR) of the valence band 
	maxima (VBM) of C-dia in both the approximations reported by Ponc{\'e} et 
	al. \cite{Ponce2015} and Tutchton et al. \cite{Tutchton2018}. The ZPR of 
	VBM is 155 meV and 134 meV in the adiabatic and non-adiabatic 
	approximations respectively \cite{Ponce2015}. Similar results, 145 meV and 
	133 meV respectively, for a different smearing parameter, have been 
	reported by Tutchton et al. \cite{Tutchton2018}. \textit{These values 
	indicate that the EPI contributions to the band-structure energies,  
	$\Delta E^{\textrm{ep}}_{\textrm{bs,adia}}$ and $\Delta 
	E^{\textrm{ep}}_{\textrm{bs,non-adia}}$, will also be different.}

	In our earlier work \cite{Varma2022}, we reported the  $\Delta 
	E^{\textrm{ep}}_{\textrm{bs,adia}}$ for C-dia to be -47 meV/atom in the 
	adiabatic approximation. We had also calculated $\Delta 
	E^{\textrm{ep}}_{\textrm{bs,non-adia}}$ to be -29 meV/atom. We have 
	computed the ZPVE of C-dia as 179 meV/atom.

	From  Eq. \ref{Eph-other}, we obtain $E^{\textrm{ph,other}}$ to be 273 
	meV/atom in the adiabatic approximation and to be 237 meV/atom in the 
	non-adiabatic approximation, i.e., $E^{\textrm{ph,other}}$ varies by 36 
	meV/atom instead of being a constant.
	
	 We now present results for Silicon Carbide-cubic (IR-active) reported 
	 earlier \cite{Varma2022} for the non-adiabatic approximation. We had also 
	 computed the results in the adiabatic approximation.

	We note that divergence issues emerge for IR-active materials when a small 
	 smearing parameter (1-10 meV) and very dense q-point grids are used 
	 \cite{Ponce2015}.  We had used a smearing 	parameter of 100 meV and a 
	 relatively less dense q-point grid so that 	divergence issues are not 
	 relevant \cite{Ponce2015}.

	The  $\Delta E^{\textrm{ep}}_{\textrm{bs,non-adia}}$ was +108 meV/f.u 
	\cite{Varma2022} and $\Delta E^{\textrm{ep}}_{\textrm{bs,adia}}$ was +64 
	meV/f.u. The ZPVE value has been reported to be 220 meV/f.u. in Ref. 
	\cite{Karch1996}. Therefore, applying Eq. \ref{Eph-other} above, we 
	obtain 	$E^{\textrm{ph,other}}$ to be 4 meV/f.u in the non-adiabatic 
	approximation and 92 meV/f.u. in the adiabatic approximation. 

	The large variation in $E^{\textrm{ph,other}}$ is unacceptable because it 
	must be a constant, as discussed above. Therefore, these numerical results 
	support our conclusion in point-2 that $\Delta 
	E^{\textrm{ep}}_{\textrm{bs,non-adia}}$ is not a part of Eq. 
	\ref{Eph-total} (authors' Eq.112). Hence, the authors' key result that 
	$E^{\textrm{Allen-EPI}}$ is part of the phonon energy, ZPVE, is not valid 
	in the non-adiabatic approximation.

\subsection*{Point-4}
\subsubsection*{$E^{\textrm{Allen-EPI}}$, in the adiabatic approximation is 
`theoretically unjustifiable' due to Allen's objections}

In the previous discussions, we have emphasized that  Eq. \ref{Eph-total} 
(authors' Eq.112), has been derived for the special case of the adiabatic 
approximation and is not valid for the general case of the non-adiabatic 
approximation.

We now discuss that $E^{\textrm{Allen-EPI}}$ is theoretically invalid even in 
the adiabatic approximation for an entirely different reason, viz. on 
thermodynamic grounds, which follows from observations of Allen 
\cite{Allen2022EPIBS}.

Allen has discussed the free energies of electrons and phonons in the 
Quasi-Harmonic (QH) Approximation  and beyond (including Quasiparticle (QP) 
interactions) under non-interacting and interacting conditions, 
respectively \cite{Allen2020}.

	The starting point is the eigenenergy for electrons given by Eq.1 of 
	Allen \cite{Allen2020} as
\begin{equation}
	\epsilon_{K}^{\mathrm{QP}}(V,T) =\epsilon_{K}(V_{0}) + 
	\Delta\epsilon_{K}^{\mathrm{QH}} + \Delta\epsilon_{K}^{\mathrm{QP}}(V,T)
	\label{Allen-Eq.1-2020}
\end{equation}
 and the eigenenergy for phonons given by Eq.2 of Allen \cite{Allen2020} as
\begin{equation}
	\omega_{Q}^{\mathrm{QP}}(V,T) = \omega_{Q}(V_{0}) + 
	\Delta\omega_{Q}^{\mathrm{QH}}	+ \Delta\omega_{Q}^{\mathrm{QP}}(V,T)
	\label{Allen-Eq.2-2020}
\end{equation}	
We refer to Allen \cite{Allen2020} for the meaning of the various symbols and 
terms.

Briefly, if the eigenenergies of electrons and phonons are given only by the 
first two terms in Eq. \ref{Allen-Eq.1-2020} and Eq. \ref{Allen-Eq.2-2020}, 
they correspond to that of non-interacting electrons and phonons 
\cite{Allen2020} respectively. In this case, the QH free energies are generated 
from the QH  energies and given by Eq.10 of Allen \cite{Allen2020} as
\begin{equation}
	F_{QH} = -k_B T  ln(Z_{el}^{bare} \times Z_{ph}^{bare})  = 	F_{el}^{QH} 
	\times 	F_{ph}^{QH}
	\label{Allen-Eq.10-2020}
\end{equation}	

The free energies for non-interacting electrons and phonons are given by Eq.11 
and Eq.12 of Allen respectively \cite{Allen2020}. In particular, the free 
energy of non-interacting electrons is given as \cite{Allen2020}

\begin{equation}
\begin{aligned}
&F_{\mathrm{el}}^{\mathrm{QH}}(V,T)
	= k_B T \sum_{K} \ln\!\left(1-f_{K0}(V,T) \right) \\
			&= \sum_{K}^{\mathrm{occ}} \left(\epsilon_{K}(V)-\mu(V)\right) + 
			k_B T \sum_{K}  \ln\!\left( \frac{1-f_{K0}(V,T)} {1-f_{K0}(V,0)}  
			\right) \\
			&\rightarrow E_{\mathrm{el}}(V) + k_B T \sum_{K} \ln\!\left(
			\frac{1-f_{K0}(V,T)} {1-f_{K0}(V,0)} \right) \\
			&\equiv E_{\mathrm{el}}(V) + \Delta 
			F_{\mathrm{el}}^{\mathrm{QH}}(V,T).
\end{aligned}
\label{Allen-Eq.11-2020}
\end{equation}	
In this seminal paper \cite{Allen2020}, Allen  has also stated that `\textit{It 
is unfortunately incorrect to insert the full renormalized energies 
into the non-interacting free energy $F_{QH}$.}" That is, if Eq. 
\ref{Allen-Eq.1-2020} also includes the contribution from the third term 
($\Delta\epsilon_{K}^{\mathrm{QP}}$), the free energy formula for 
non-interacting electrons (Eq. \ref{Allen-Eq.11-2020}) cannot be used.  

In our earlier work \cite{Varma2022}, we had approximated Allen's equation, 
Eq. \ref{EPI-zero-K-Allen}, to be the EPI contribution to the band-structure 
energy, i.e. $\Delta E^{\textrm{ep}}_{\textrm{bs}}$. In his named review 
\cite{Allen2022EPIBS} of earlier version of that paper \cite{Varma2022}, Allen 
objected that $\Delta E^{\textrm{ep}}_{\textrm{bs}}$ as the \textit{sole term} 
in the EPI contribution is not theoretically justifiable, as it can be derived 
from the free energy expression for non-interacting particles. 

The following derivation is reproduced from Allen \cite{Allen2022EPIBS}. 
``The free energy shift caused by  the shift in eigenenergies due to the EPI, 
$\Delta \epsilon _K^{\textrm{ep}}$, is,
\begin{equation}
\begin{split}
	\Delta F_{\text{epi}} =& k_B T\sum_{K}  \ln\left[\frac{1 - 
			f(\epsilon_K^{QH} + \Delta \epsilon_K^{\textrm{ep}})}{1 - 
			f(\epsilon_K^{QH})}\right] \\      
    \approx & \sum_{K} f_K^{QH} \Delta 
	\epsilon_K^{\textrm{ep}} \rightarrow \Delta 
		E^{\textrm{ep}}_{\textrm{bs}}
	    \label{EPI-non-int-1}
        \end{split}
        \end{equation}
 	when expanded to the lowest order \cite{Allen2022EPIBS}". Allen further 
 	commented that ``The method used here (putting 	band-electron energies, 
 	$\epsilon_K$, including electron-phonon corrections, 
 	$\epsilon_K^{\textrm{ep}}$, into the non-interacting electron free 
 	energy formula) is not theoretically justifiable." 	\cite{Allen2022EPIBS}.

\textit{The same issue has emerged in the context of authors' key result and is 
discussed below.}

The authors have shown that, in the adiabatic approximation, Allen's equation 
for the EPI contribution to the total energy becomes the EPI contribution to 
the band-structure energy, $E^{\textrm{Allen-EPI}}$, their Eq.107 
\cite{Ponce2026PRXEnergy}. (This is also discussed in Point-1 where 
$E^{\textrm{Allen-EPI}}$ is referred to as  $\Delta 
E^{\textrm{ep}}_{\textrm{bs,adia}}$ for clarity.)

Because $\Delta E^{\textrm{ep}}_{\textrm{bs,adia}}$  is the \textit{sole} EPI 
contribution to the total energy, Allen's objections \cite{Allen2020, 
Allen2022EPIBS} discusssed above are applicable and hence, this result is 
theoretically unjustifiable.  Because the authors' key result, Eq.112,  (Eq. 
\ref{Eph-total} above), contains the $\Delta 
E^{\textrm{ep}}_{\textrm{bs,adia}}$ term, it follows that the authors' Eq.112 
is also not theoretically justifiable due to Allen's objections.

	Thus, only the EPI contribution to the total energy in the non-adiabatic 
	approximation, Eq. \ref{EPI-zero-K-Allen-non-adia}, is theoretically valid 
	for all semiconductors and insulators, because the second term ($\Delta 
	E_{\textrm{tot,non-adia}}^{\textrm{FM,occ}}$) is non-zero. It follows that 
	Eq. \ref{EPI-zero-K-Allen-non-adia} cannot be derived from the free energy 
	formula for non-interacting electrons. Hence, Allen's objections do not 
	apply to Eq. \ref{EPI-zero-K-Allen-non-adia}.

\subsection*{Point-5}
\subsubsection*{The result that  $E^{\textrm{Allen-EPI}}$ is a part of ZPVE is 
a matter of interpretation and not  uniquely established}

	Allen's objections, discussed above in Point-4, raise questions about 
	 Eq. \ref{Eph-total} (authors' Eq.112). This motivated us to re-examine  
	 authors' Eq.112 \cite{Ponce2026PRXEnergy}. 
	
	It is well known that the DFT total energy can be written in two forms 
	\cite{Giustino2014materials}. These two expressions are given in the 
	authors' \cite{Ponce2026PRXEnergy} Eq. 88 and Eq.90.  
	
	We refer to the detailed discussion in Exercise 9.1 of Giustino	
	\cite{Giustino2014materials}. In the usual form, the total electronic 
	energy is given as \cite{Giustino2014materials} 
	\begin{equation}
	\begin{split}
	E =& \\ 	\sum_i f_i &\int dr\, \phi_i^*(\textbf{r})
	\left[ -\frac{\nabla^2}{2} + V_n(\textbf{r}) + \frac{1}{2}
	\int dr'\, \frac{n(\textbf{r}')}{|\textbf{r}-\textbf{r}'|} \right] 
	\phi_i(\textbf{r})\\ +& E_{\mathrm{xc}}[n]
	\label{Giustino tot en usual}
    \end{split}
    \end{equation}

	The total energy includes the contribution from the nuclear-nuclear 
	repulsion in addition to the electronic energy given above and is given by 
	Eq. 2.45 of Giustino \cite{Giustino2014materials}, i.e., $E_{tot} = E + 
	E_{NN}$
		
	To obtain an alternate form, which contains the band-structure energy, the 
	quantity $\left[V_H/2 + V_{xc}\right]$ is added and subtracted 
	\cite{Giustino2014materials}. When added, it leads to 
	the eigenvalue equation and the total energy can be rewritten as 
	\cite{Giustino2014materials}
	\begin{equation}
	\begin{split}
	E_{tot} =& \sum_i f_i \int d\textbf{r}\, \phi_i^*(\textbf{r})
	\left[ -\frac{\nabla^2}{2} + V_{\mathrm{tot}}(\textbf{r}) \right]
	\phi_i(\textbf{r}) \\ &- \int d\textbf{r}\, n(\textbf{r})
	\left[ \frac{1}{2}V_{\mathrm{H}}(\textbf{r}) + V_{\mathrm{xc}}(\textbf{r}) 
	\right]
	+ E_{\mathrm{xc}}[n] + E_{NN}
	\label{Giustino totEn alternate}
	\end{split}
	\end{equation}
	The first line of Eq. \ref{Giustino totEn alternate} corresponds 
	to the authors' Eq.92 for $E^{\textrm{e}}\{\textbf{R}\}$ and the second 
	line corresponds to the authors' Eq.93 for 
	$E^{\textrm{other}}\{\textbf{R}\}$ 	\cite{Ponce2026PRXEnergy} (where the 
	quantity $\left[V_H/2 + V_{xc}\right]$ is  $V^\textrm{Hxc}$).

	We highlight the fact that the 	quantity  $\left[V_H/2 + V_{xc}\right]$ has 
	been artificially added and subtracted. When added it becomes a part of 
	$E^{\textrm{e}}\{\textbf{R}\}$ and eventually (through Eq.109) contributes 
	to EPI part of ZPVE. When subtracted it becomes a part of 
	$E^{\textrm{other}}\{\textbf{R}\}$ and eventually (through Eq.113) 
	contributes to the $E^{\textrm{ph,other}}$ part of ZPVE. This dual role for 
	an artificially introduced term is questionable and is a matter of 
	interpretation and not uniquely justifiable.

	If this quantity is added and eventually becomes a part of EPI 
	contribution, then when subtracted also, it should eventually be a part of 
	the EPI contribution, because in both cases, the second derivative is taken 
	(authors' Eq.109 and Eq.113). That is, the $E^{\textrm{ph,other}}$ term 
	will include an EPI contribution due to the quantity $\left[V_H/2 + 	
	V_{xc}\right]$. Because the quantity $\left[V_H/2 + V_{xc}\right]$ has 
	opposite signs in the two terms, the net effect is that its contribution is 
	cancelled, and hence, $E^{\textrm{Allen-EPI}}$ is not part of ZPVE. 
	Therefore,  we submit that Eq. \ref{Eph-total} (authors' Eq.112) is not 
	uniquely established but is a matter of interpretation.

\section{Computational concerns}

\subsection*{Point-6}
\subsubsection*{The new formalism cannot account for the experimental energy 
differences between Boron Nitride polymorphs}
In the computational studies of Boron Nitride (BN) polymorphs, the stable 
polymorph  has been controversial for decades. Experimental (DTA) studies 
clearly 
establish that cubic-BN (c-BN) is the stable polymorph 
\cite{Solozhenko1995boron,Gavrichev1993low}. Further, the hexagonal-BN (h-BN) 
is metastable by  $>$  150 meV/f.u. 
\cite{Solozhenko1995boron,Gavrichev1993low}. These results are discussed in 
detail elsewhere \cite{Paul2022}. In computational studies, DFT based studies 
have included the DFT total energy, the dispersion (vdW) contribution and the 
ZPVE contribution to the total energy \cite{Cazorla2019, Nikaido2022diffusion}. 
The dispersion corrections include the semi-empirical DFT-D methods 
\cite{Cazorla2019}  and also more advanced methods e.g.  the  diffusion monte 
carlo (DMC)\cite{Nikaido2022diffusion}, random phase approximation (RPA) 
\cite{Cazorla2019}. The stability order was sensitive to the dispersion 
approximation used, with either the \textrm{sp$^3$-bonded} cubic (BN-3C) or the 
\textrm{sp$^2$-bonded} hexagonal (h-BN) as the stable polymorph 
\cite{Cazorla2019}.  Importantly, the above computational studies reported 
energy differences of  $\sim$ 30 meV/f.u and none could  match the experimental 
energy differences of $>$ 150 meV/f.u 
\cite{Solozhenko1995boron,Gavrichev1993low}. In summary, even the most advanced 
computational methods are inconsistent in the prediction of the stable BN 
polymorph. Further, the predictions don't match the experimental results. 

\textit{We have highlighted the above inconsistencies because the total energy 
in the new formalism, upto second-order, is identical to that used in the above 
studies, i.e. DFT (+vdW) + ZPVE (Table III of Ref. 
\cite{Ponce2026PRXEnergy})}.   Hence, it is clear that if the total energy 
expression in the new formalism were to be used, the controversy about the 
stable BN polymorph will persist without any resolution.

However, in the effective Hamiltonian formalism, the EPI term must be included 
separately \cite{Allen2022erratum}. Tutchton et al. report the ZPR of VBM of 
c-BN is 133 meV and h-BN is 253 meV \cite{Tutchton2018}. This is a clear 
indication that the EPI contribution to the total energy will also be 
significantly different for c-BN and h-BN. This is indeed the case and our 
results \cite{Paul2022} show that c-BN is the stable polymorph independent of 
the DFT-D approximation. Further, h-BN is metastable by $>$ 150 meV/f.u. 
compared to c-BN. Hence, \textit{for the first time, by including the EPI 
contribution as a separate term, it becomes possible to account for the 
experimental results \cite{Paul2022}}. 

The controversy regarding the stable BN polymorph provides a stringent test of 
the total energy expression in the two formalisms. Our results \cite{Paul2022} 
clearly suggest that the total energy in the effective Hamiltonian formalism is 
to be preferred, as it can give a better match with the experimental results.

\subsection*{Point-7}
\subsubsection*{The results on GaN are theoretically unjustified at present}

In their Table IV, the authors \cite{Ponce2026PRXEnergy} discuss the total 
energy and the relative stability of GaN polymorphs. The authors explicitly 
state ``we are reporting the nonadiabatic results since it is a polar material".

As discussed above, \textit{Eq. \ref{Eph-total} (authors' Eq.112) is not 
theoretically justifiable in the non-adiabatic approximation}. Therefore, the 
GaN results are not justified at present and must be kept in abeyance until the 
concerns raised in the above points are resolved.

\section{Summary}

	The authors have proposed a new formalism for the EPI contribution to the 
	total energy. A key result, obtained in the adiabatic approximation, is 
	that the EPI contribution to the total energy (as derived by Allen) is 
	alrealy included in the phonon energy. We show that this result  is not 
	theoretically justifiable in the general case of the non-adiabatic 
	approximation, which must be used for all IR-active materials. In addition, 
	using Allen's objections, we show that the key result is not 
	theoretically justifiable even in the specific case of the adiabatic 
	approximation. For BN polymorphs, we show that the total energy in the 
	effective Hamiltonian formalism gives a better match to experimental 
	results. For all of the above reasons the effective Hamiltonian formalism 
	must be preferred in total energy studies.

\section{Acknowledgement}
We thank Prof. M.P. Gururajan for helpful discussions.

\bibliography{references.bib}

\end{document}